\documentclass[conference]{IEEEtran}
\usepackage{pifont}
\usepackage{graphicx}
\usepackage[cmex10]{amsmath}
\usepackage{epsfig}
\usepackage{bm}
\usepackage[usenames]{color}
\usepackage{multirow}
\usepackage{array}
\usepackage{amssymb}
\usepackage{cite}
\usepackage{notoccite} 
\usepackage{multirow}
\usepackage{color, soul}                        
\usepackage{epstopdf}
\usepackage{enumerate}
\usepackage{caption}
\usepackage{subcaption}
\usepackage{amscd}
\usepackage{amsmath}

\usepackage{bbm}

\ifCLASSINFOpdf
  
\else
 
\fi

\begin{document}
%
\title{Overlay Secondary Spectrum Sharing with Independent Re-attempts in Cognitive Radios}



%
\author{\IEEEauthorblockN{Muthukrishnan Senthil Kumar\IEEEauthorrefmark{1},
Aresh Dadlani\IEEEauthorrefmark{2},
Kiseon Kim\IEEEauthorrefmark{2} and
Richard O. Afolabi\IEEEauthorrefmark{3}}
\IEEEauthorblockA{\IEEEauthorrefmark{1}Department of Applied Mathematics and Computational Science,
PSG College of Technology, India}
\IEEEauthorblockA{\IEEEauthorrefmark{2}School of Electrical Engineering and Computer Science,\\
Gwangju Institute of Science and Technology, Gwangju 61005, South Korea}
\IEEEauthorblockA{\IEEEauthorrefmark{3}Computer Science Department, University of Nevada - Las Vegas, NV 89119, USA\\
E-mail: msk@amc.psgtech.ac.in, \{dadlani, kskim\}@gist.ac.kr, ao.richard@me.com}
}


\maketitle

\begin{abstract}
Opportunistic spectrum access (OSA) is a promising reform paradigm envisioned to address the issue of spectrum scarcity in cognitive radio networks (CRNs). While current models consider various aspects of the OSA scheme, the impact of retrial phenomenon in  multi-channel CRNs has not yet been analyzed. In this work, we present a continuous-time Markov chain (CTMC) model in which the blocked/preempted secondary users (SUs) enter a finite retrial group (or orbit)  and re-attempt independently for service in an exponentially distributed random manner. Taking into account the inherent retrial tendency of SUs, we numerically assess the performance of the proposed scheme in terms of dropping probability and throughput of SUs. 
\end{abstract}

\IEEEpeerreviewmaketitle

\section{Introduction}
\IEEEPARstart{P}{roliferation} of wireless services in recent years has stimulated the need for improved spectrum management policies to exploit the abundant spectrum holes in existing licensed bands. Cognitive radio (CR) has emerged as a potential technology that allows unlicensed or secondary users (SUs) to temporarily access idle licensed bands exclusive to primary users (PUs) \cite{Haykin}. To avoid interference with PUs, SUs continuously sense the spectrum using artificial intelligent techniques and are preempted when a newly arriving PU finds insufficient bandwidth for its transmission. Therefore, the degree of flexibility in spectrum sharing between PUs and SUs is regulated by the opportunistic spectrum access (OSA) policy in use \cite{Min}\cite{Yang}.

Modelling OSA schemes for CRNs has been the focus of many recent active research studies. In \cite{Thi}, an OSA model with a single primary traffic for the PUs and two prioritized secondary traffic classes for the SUs was studied. In this work, the authors assumed that the blocked and preempted SUs had no option other than to quit the system in absence of vacant bands. In practical scenarios however, such SUs are expected to retry for service after some random amount of time. With regard to the retrial behavior of SUs, the authors of \cite{Zhu} proposed a continuous-time Markov chain (CTMC) with level-dependent quasi-birth-and-death (QBD) structure to analyze the effect of retrying SUs on the performance of an unslotted CRN. Nonetheless, they dealt with the carrier sense multiple access (CSMA) protocol in a mixed environment of dedicated licensed and unlicensed bands. In \cite{Chang}, an $M/M/1$ retrial queueing framework for OSA was proposed for single channel CRNs. In this paper, we divulge the impact of the retrial behavior of opportunistic SUs in multi-channel CRNs where all spectrum bands are licensed to PUs. We express the state transitions of the proposed OSA model as a tri-variate CTMC and investigate the system performance in terms of the dropping probability and network throughput of SUs.

The rest of this paper is structured as follows. Section~\ref{sec2} describes and formulates the system model, followed by derivations of the performance measures in Section~\ref{sec3}. Analytical and simulation results are extensively discussed in Section~\ref{sec4}. Finally, conclusive remarks are drawn in Section~\ref{sec5}.
\begin{figure}[!t]
	\centering
	\includegraphics[width=3.1in]{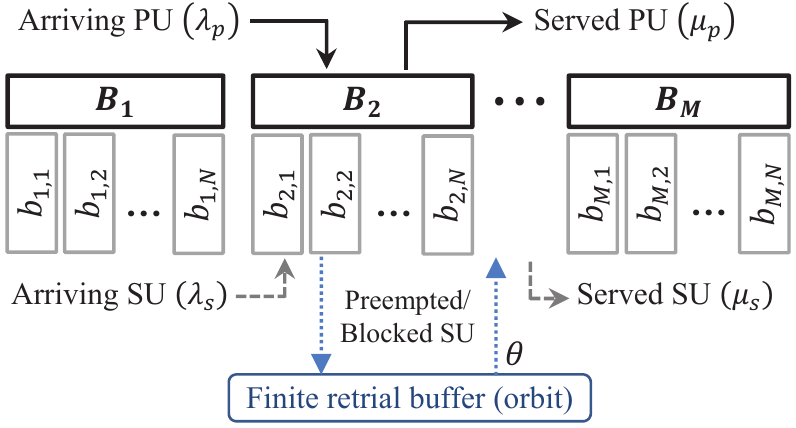}
	\caption{The proposed OSA model with retrial phenomenon.}
	\label{fig1}
	\vspace{-0.5em}
\end{figure}
%

\section{System Model Formulation}
\label{sec2}
We consider a multi-channel CRN comprising of $M$ licensed bands, each divided into $N$ sub-bands. Henceforth, we use the terms `band' and `sub-band' to refer to a bandwidth unit for PU and SU traffic sources, respectively. As shown in \figurename{~\ref{fig1}}, a PU requires one band $(B_i)$ while an SU needs one sub-band $(b_{i,j})$ for transmission. Hence, when a given band is exploited by a PU, the underlying sub-bands are made unavailable to SUs. A newly arriving PU occupies a band if vacant. Otherwise, it preempts the SUs using the sub-bands of one of the primary bands. The preempted SUs then enter the orbit of size $L$ and retry again after some random time with rate $\theta$. On retrial, if a sub-band is idle, the SU opportunistically utilizes it. Otherwise, it repeats the cycle. A preempted or blocked SU is dropped only if the orbit is full. We assume that PU and SU arrivals follow Poisson processes with rates $\lambda_p$ and $\lambda_s$, respectively. Moreover, let $\mu_p$ and $\mu_s$ denote the mean service rates of the exponentially distributed service times of PU and SU, respectively. Let $N_p(t)$ and $N_s(t)$ be the number of bands and sub-bands used by PUs and SUs, respectively, at time $t$. Also, let $N_r(t)$ denote the number of SUs in the orbit at time $t$. Hence, $\{(N_p(t),N_s(t),N_r(t))|t \geq 0\}$ is a three-dimensional CTMC with state space $\Omega=\{(i,j,k)|i=0, \ldots, M; j=0, \ldots, \min(MN,(M-i)N); k=0, \ldots, L\}.$ Assuming the elements of $\Omega$ to be ordered lexicographically, the infinitesimal generator $Q$ of the above process is a finite level-dependent QBD type (LDQBD) given as follows:
\begin{equation}
Q =
 \begin{pmatrix}
  A_0		& D_1		& 0			& 0			& \cdots & 0		& 0 		\\
  C_1		& A_1		& D_2		& 0			& \cdots & 0		& 0			\\
  0			& C_2		& A_2		& D_3		& \cdots & 0		& 0			\\
  0			& 0			& C_3		& A_3		& \cdots & 0		& 0			\\
  \vdots	& \vdots	& \vdots	& \ddots	& \ddots & \vdots	& \vdots	\\
  0			& 0			& 0			& 0			& \cdots & A_{M-1}	& D_M		\\
  0			& 0			& 0			& 0			& \cdots & C_M		& A_M
 \end{pmatrix}\,,
\end{equation}
where the block matrices $A_i$, $D_i$, and $C_i$ are of order $X(L+1)$, $X(L+1) \times (X-N)(L+1)$, and $(X+N)(L+1)\times X(L+1)$, respectively, with $X$ defined as $(M-i)N+1$. Matrix $A_i$ is the transition rate matrix in level $i$ with the following structure:
\begin{equation}
\label{A_i}
A_i =
 \begin{pmatrix}
  A_{0,0}	& A_{0,1}	& 0			& \cdots & 0			& 0				\\
  A_{1,0}	& A_{1,1}	& A_{1,2}	& \cdots & 0			& 0				\\
  0			& A_{2,1}	& A_{2,2}	& \cdots & 0			& 0				\\
  0			& 0			& A_{3,2}	& \cdots & 0			& 0				\\
  \vdots	& \vdots	& \vdots	& \ddots & \vdots		& \vdots		\\
  0			& 0			& 0			& \cdots & A_{X-2,X-2}	& A_{X-2,X-1}	\\
  0			& 0			& 0			& \cdots & A_{X-1,X-2}	& A_{X-1,X-1}
 \end{pmatrix}.
\end{equation}
In (\ref{A_i}), $A_{j,j-1}$ is the transition rate sub-matrix for SU service completion and $A_{j,j+1}$ indicates the service/retrial requests of SUs. With the Kronecker delta function and identity matrix of order $Y$ denoted as $\delta_{i,j}$ and $I_Y$, respectively, the elements of $A_i$ are given as below, where each element is a square matrix: 
\begin{equation}
	\left\{
	\begin{aligned}
		&\!A_{j,j}\!=\!-[\lambda_s\!+\!\big(1\!-\!\delta_{i,M}\big)\big(\lambda_p\!+\!(1\!-\!\delta_{j,MN})k\theta\big)]I_{L+1}\\
		&\qquad\qquad\qquad\qquad\qquad\qquad\qquad\!-\!j\mu_s I_{L+1}\!-\!i \mu_p I_{L+1},\\
		&\!A_{j,j-1}\!=\!j \mu_s I_{L+1},\\
		&\!A_{j,j+1}\!=\!\begin{pmatrix}
  						\lambda_s	& 0			&  0		& \cdots & 0 			& 0			\\
  						\theta		& \lambda_s	& 0			& \cdots & 0 			& 0			\\
   						0			& 2\theta	& \lambda_s	& \cdots & 0 			& 0			\\
   						0			& 0			& 3\theta			& \cdots & 0 			& 0			\\
 						\vdots		& \vdots	& \vdots	& \ddots & \vdots		& \vdots	\\
    					0			& 0         & 0			& \cdots & \lambda_s 	& 0			\\
   						0			& 0         & 0         & \cdots & L\theta 		& \lambda_s
 					\end{pmatrix}\,,\\
		&\!A_{X-1,X-1}\!=\!-[\lambda_s\!+\!(1\!-\!\delta_{i,M})\lambda_p\!+\!(X\!-\!1)\mu_s] I_{L+1}\\
		&\qquad\qquad\qquad\qquad\qquad\qquad\qquad\qquad\!-\!i \mu_p I_{L+1}\!+\!\lambda_s Z,
	\end{aligned}
	\right.
\end{equation}
and $Z = [\textbf{0}|I_L|\textbf{0}^T]$ is an augmented matrix of order $L+1$. Transitions associated with PU service completion from level $i$ to $(i-1)$ are given in sub-matrix $C_i=i \mu_p [I_{X(L+1)}|\textbf{0}]$. The block matrix $D_i$ holds the transition rates from level $(i-1)$ to $i$ for PU requests. The rate from state $(i-1,j,k)$ to state $(i,j,k)$ is $\lambda_p$ if $j\leq X\!+\!N$. The transition from state $(i-1,j,k)$ to $(i,j-l,k+l)$ occurs with rate $\lambda_p$ if $j > X\!+\!N$, $l=1,2,\ldots,N$, and $k \leq L\!-\!1$. The loss of SUs due to PU preemption from state $(i\!-\!1,j,L)$ to $(i,j\!-\!l,L)$ takes place at rate $\lambda_p$ for $l=1,2,\ldots,N$ and $j\! >\! X\!+\!N$. Now, let $\Pi$ be the steady-state probability vector satisfying equations $\Pi Q=0$ and $\Pi e=1$, where $e$ represents a unit column vector. Furthermore, we partition $\Pi$ into $\left(\Pi_0,\Pi_1,\ldots,\Pi_M\right)$, where vector $\Pi_i=\left(\Pi_{i,0},\Pi_{i,1},\ldots,\Pi_{i,\min(MN,(M-i)N)}\right)$ and vector $\Pi_{i,j}=\left(\pi_{i,j,0},\pi_{i,j,1},\ldots,\pi_{i,j,L}\right)$. Thus, $\pi_{i,j,k}$ is the steady-state probability that the system is in state $(i,j,k)$. The $m^{th}$ $\pi_{i,j,k}$ element of $\Pi$ can be acquired as follows:
\begin{equation}
m = \left[\sum_{l=1}^{i}((M\!-\!i\!+\!1)N\!+\!1)\right](L\!+\!1)\!+\!j(L\!+\!1)\!+\!(k\!+\!1). 
\end{equation}
Undertaking the same approach of \cite{Essia}, $\Pi_i$ can be obtained in following matrix geometric form, where $R_M\!=\!-D_M A_M^{-1}$ and $R_i\!=\!-D_i [A_i\!+\!R_{i+1}C_{i+1}]^{-1}$ for $1 \leq i \leq M\!-\!1$:
\begin{equation}
\label{main_eq}
\Pi_i=\Pi_1 \prod_{j=1}^{i} R_j ;\qquad  i=1,2,\ldots,M,
\end{equation}
with the normalization $\Pi_1 \big(R_0 e_0\!+\!\sum_{i=1}^{M} \prod_{j=1}^{i} R_j e_i\big)\!=\! 1$, where $e_i$ denotes the unit column vector.

\section{Performance Measures}
\label{sec3}
In this section, we derive the dropping probability $(P^{SU}_{drop})$ and throughput $(T^{SU})$ of SUs using the steady-state distribution obtained earlier. An SU is said to be dropped from the system in any one of the following mutually-exclusive events:
\begin{itemize}
	\item When an SU finds all bands occupied on arrival and the retrial orbit full, it gives up its service with rate $\lambda_s \pi_{M,0,L}$.
	\item When an arriving SU finds all sub-bands already taken by other SUs and the orbit full, it leaves with rate $\lambda_s \sum_{i=0}^{M-1} \pi_{i,(M-i)N,L}$.
	\item When an SU in service is preempted by an arriving PU and finds the orbit full, it quits the system with rate $\lambda_p \sum_{i=0}^{M}\sum_{l=1}^{N} \pi_{i,(M-i)N-l,L}$.
\end{itemize}
Consequently, $(P^{SU}_{drop})$, defined as the ratio of the dropping rate of SUs to their arrival rate, is derived to be:
\begin{equation}
\label{drop_SU}
P_{drop}^{SU} = \sum_{i=0}^{M} \pi_{i,(M-i)N,L} + \frac{\lambda_p}{\lambda_s}\left[\sum_{i=0}^{M}\sum_{l=1}^{N} \pi_{i,(M-i)N-l,L}\right].
\end{equation}

The achievable throughput of SUs is the number of SUs that succeed in accessing sub-bands per unit time and is \cite{Zhu}:
\begin{equation}
\label{t_SU}
T^{SU} = \lambda_s(1 - P_{drop}^{SU}).
\end{equation}

\section{Numerical and Simulation Results}
\label{sec4}
We merely focus on the impact of SU retrials on the system performance in this section. Without loss of generality, we set $\lambda_s\!=\!1.5$, $\lambda_p\!=\!0.1$, $\mu_s\!=\!0.4$, and $\mu_p\!=\!0.2$ in all scenarios. Also, all simulation results have been averaged over 100 runs.

\begin{figure}[!t]
	\centering
	\includegraphics[width=3.5in]{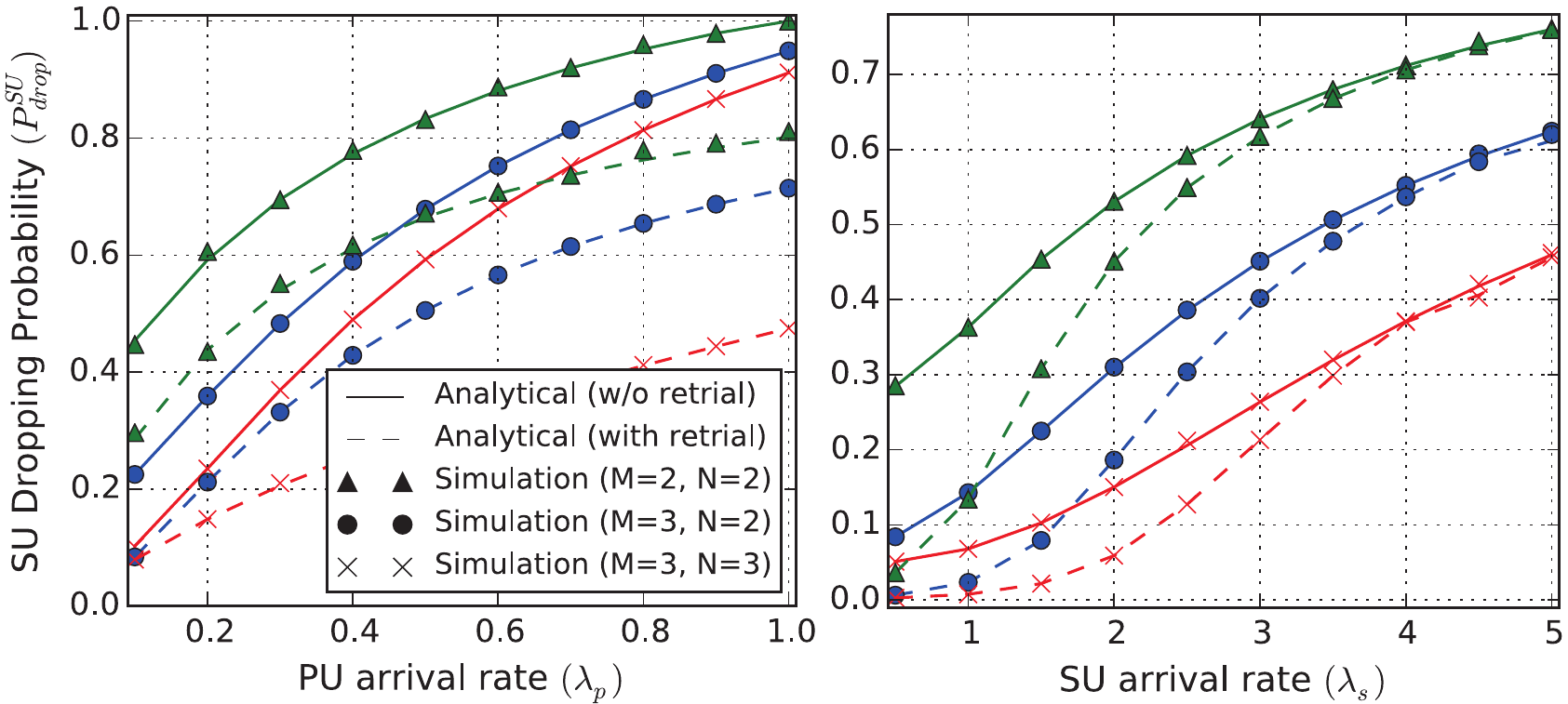}
	\caption{SU dropping probability with respect to arrival rates.}
	\label{fig2}
	\vspace{-0.5em}
\end{figure}

\figurename{~\ref{fig2}} depicts $P^{SU}_{drop}$ as functions of the user arrival rates for varying $(M,N)$ values with $L\!=\!10$ and $\theta\!=\!2$. To highlight the retrial effect, our scheme is compared with a non-retrial system (i.e. $L\!=\!0$). We observe that $P_{drop}^{SU}$ increases with the arrival rate of PUs. This probability however, reduces substantially with increase in bands and sub-bands. In comparison to the non-retrial system, the proposed model yields lower $P_{drop}^{SU}$ values. For instance, at $\lambda_p\!=\!0.4$, $P_{drop}^{SU}$ reduces by almost $30\%$ as $M$ increases to 3 and further by about $40\%$ as $N$ rises to 3 in presence of retrying SUs. Similarly, $P_{drop}^{SU}$ increases with the arrival rate of SUs. However, as $\lambda_s$ increases, $P_{drop}^{SU}$ in our scheme converges to that of the non-retrial system accounting for unavailable sub-bands and full orbit size.
\begin{figure}[!t]
	\centering
	\includegraphics[width=3.5in]{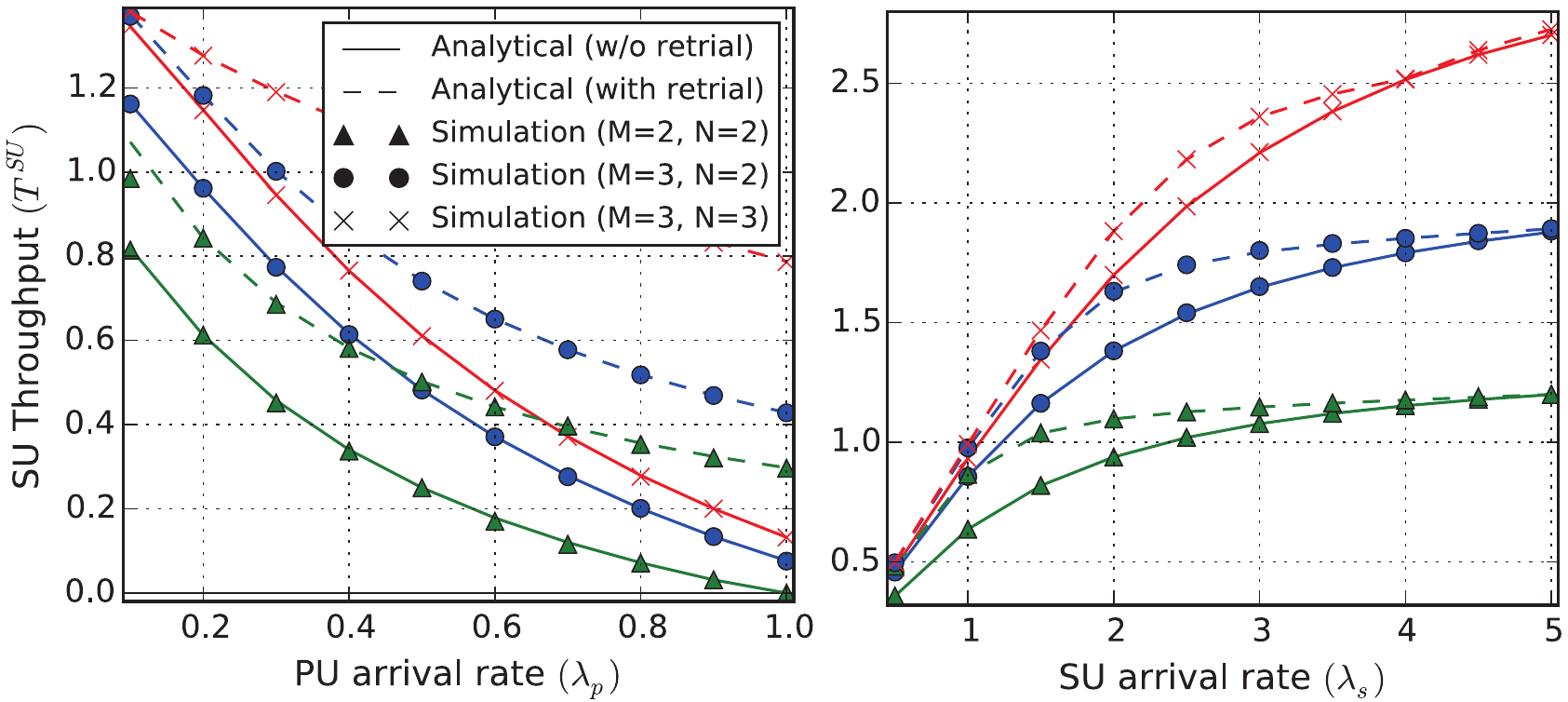}
	\caption{SU throughput with respect to arrival rates.}
	\label{fig3}
	\vspace{-0.5em}
\end{figure}

For the same system parameters, the effect of $\lambda_p$ and $\lambda_s$ on the throughput of SUs are shown in \figurename{~\ref{fig3}}. It is evident that for any given $(M,N)$, $T^{SU}$ decreases as $\lambda_p$ increases, whereas it interestingly rises with increase in $\lambda_s$. This is because of the retrial action of SUs that result in higher sub-band utilization as the number of SUs entering the system increases. For $M\!=\!3$ and $\lambda_p\!=\!0.5$, $T^{SU}$ steeply falls by almost $30\%$ as $N$ decreases to 2, while this difference is less than $0.37\%$ for $\lambda_s=0.5$. This clearly justifies the fact that more number of bands increases the spectrum utilization opportunity for SUs.

\begin{figure}[!t]
	\centering
	\includegraphics[width=3.5in]{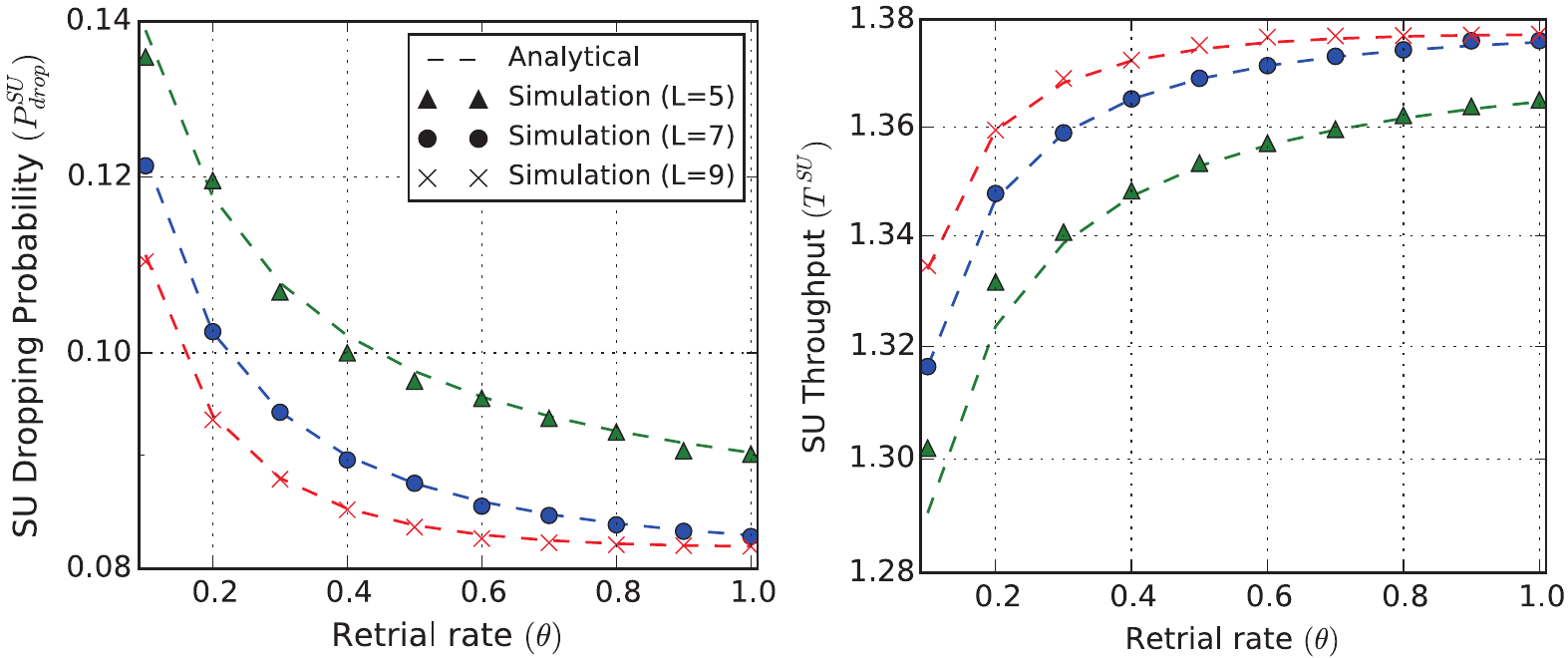}
	\caption{SU dropping probability and throughput with respect to retrial rate for varying $L$ values.}
	\label{fig4}
	\vspace{-0.5em}
\end{figure}

Finally, \figurename{~\ref{fig4}} signifies the impact of SU retrial rate on their dropping probability and throughput for $(M,N)\!=\!(3,2)$. We see that the dropping probability of SUs in the system can be reduced if the preempted/blocked SUs try to access idle sub-bands more frequently. In other words, the chances of SUs finding vacant sub-bands is higher if they keep retrying for service at higher rates. Likewise, increasing $L$ allows for the accommodation of more SUs and thus, further alleviates their probability of being forced to leave the system. The dropping probabilities for all $L$ values eventually stabilize for high retrial rates. The opposite impact however, is anticipated for SU throughput, i.e. with increase in $\theta$, more SUs are able to access the sub-bands. The higher SU throughput for orbits with larger capacities is also apparent in this figure.

\section{Conclusion}
\label{sec5}
An OSA scheme for multi-band CRNs was investigated in this paper, wherein the retrial behavior of blocked and pre-empted SUs was considered. The proposed scheme was modeled as a three-variate Markov chain and the steady-state probability distribution was obtained using matrix geometric method. Measures such as SU dropping probability and throughput were derived analytically and justified through simulation results. Results revealed the evident impact of retrial phenomenon with respect to PU and SU arrival rates, the orbit size, and retrial rate of SUs in the orbit. A possible extension to this work is considering general distributions for the inter-arrival time, service time, and retrial time.


\section*{Acknowledgment}
This research was a part of the project titled ``Development of Ocean Acoustic Echo Sounders and Hydro-Physical Properties Monitoring Systems'', funded by the ministry of Oceans and Fisheries, Korea.

\end{document}